\documentclass[aip,jcp,reprint]{revtex4-1}
\usepackage{amsmath}
\usepackage{hyperref}
\usepackage{braket}
\usepackage{color}
\usepackage[capitalise]{cleveref}
\usepackage{threeparttable}
\usepackage{graphicx}
\newcommand{\com}{}

\newcommand{\unitWm}{$[\frac{10^{33}{\rm Hz}}{{\rm e.cm}^2}]$}
\begin{document}
\title{Enhanced \texorpdfstring{$\mathcal{P,T}$}{PT}-violating nuclear magnetic quadrupole moment effects in laser-coolable molecules}
\author{Malika Denis}
\email{m.denis@rug.nl}
\affiliation{Van Swinderen Institute for Particle Physics and Gravity, Faculty of Science and Engineering, University of Groningen, 9747 AG, Groningen, The Netherlands}

\author{Yongliang Hao}
\affiliation{Van Swinderen Institute for Particle Physics and Gravity, Faculty of Science and Engineering, University of Groningen, 9747 AG, Groningen, The Netherlands}

\author{Ephraim Eliav}
\affiliation{School of Chemistry, Tel Aviv University, 69978 Tel Aviv, Israel}

\author{Nicholas R. Hutzler}
\affiliation{Division of Physics, Mathematics, and Astronomy, California Institute of Technology, Pasadena, CA 91125, USA}

\author{Malaya K. Nayak}
\affiliation{Theoretical Chemistry Section, Bhabha Atomic Research Centre, Trombay, Mumbai 400085, India}

\author{Rob G. E. Timmermans}
\affiliation{Van Swinderen Institute for Particle Physics and Gravity, Faculty of Science and Engineering, University of Groningen, 9747 AG, Groningen, The Netherlands}

\author{Anastasia Borschesvky}
\affiliation{Van Swinderen Institute for Particle Physics and Gravity, Faculty of Science and Engineering, University of Groningen, 9747 AG, Groningen, The Netherlands}

\date{\today}
\begin{abstract}
   Nuclear magnetic quadrupole moments (MQMs), like intrinsic electric dipole moments of elementary particles, violate both parity and time-reversal symmetry and therefore probe physics beyond the Standard Model of particle physics. We report on accurate relativistic coupled cluster calculations of the nuclear MQM interaction constants in BaF, YbF, BaOH, and YbOH. We elaborate on estimates of the uncertainty of our results. The implications of experiments searching for nonzero nuclear MQMs are discussed.
   
\end{abstract}

\maketitle

\section{Introduction}
One of the three conditions, first delineated by Sakharov\cite{Sakharov1967}, to explain the dominance of matter over antimatter in the Universe, is $\mathcal{CP}$ violation, the combined violation of charge conjugation (particle-antiparticle conjugation) and parity. New sources of $\mathcal{CP}$ violation, beyond those included in the Standard Model (SM) of particle physics, are required to be consistent with cosmological observations. Since in speculative extensions of the SM the $\mathcal{CPT}$ theorem is generally valid, new $\mathcal{CP}$ violation gives rise to new interactions that violate time reversal $\mathcal{T}$. Such $\mathcal{T}$-odd interactions can be observed in ordinary matter via tiny atomic or molecular energy shifts or symmetry forbidden transitions. Hence, the search for nonzero $\mathcal{T}$-odd interactions allows us to probe high-energy physics with small-scale experiments at low energy.     

In the low-energy regime $\mathcal{CP}$ violation was, and is, mostly investigated through the search for permanent electric dipole moments (EDMs) of particles, atoms, or molecules, which violate both $\mathcal{P}$ and $\mathcal{T}$. The present bounds on the EDMs of the electron and the neutron, for instance, are\cite{andreev2018improved} $|d_e|< 1.1\times10^{-29}$ e.cm  and\cite{Pendlebury2015} $|d_n|< 3.0\times10^{-26}$ e.cm, respectively. In particular, the tremendous progress in the manipulation of atoms and molecules have made it possible to dramatically lower the upper limit on the EDM of the electron in recent years. 

The EDM is not the only electromagnetic moment that violates $\mathcal{P}$ and $\mathcal{T}$. Particles or composite systems (nuclei, atoms, or molecules) with spin equal to (or larger than) 1 are expected to possess also a $\mathcal{P,T}$-odd magnetic quadrupole moment (MQM). Compared to EDMs, MQMs can be sensitive to different microscopic sources of $\mathcal{CP}$ violation\cite{Liu2012}. The search for nonzero MQMs therefore probes physics beyond the SM in a complimentary way to EDMs. Especially promising are nuclear MQMs, because in heavy deformed nuclei they can be significantly enhanced due to collective effects\cite{Sushkov1984,Ginges2004}. An example is the nucleus $^{173}$Yb, a stable isotope with spin $\geq1$. At the quark-gluon level, nMQMs can arise from the QCD vacuum angle and from dimension-six operators that originate from physics beyond the SM\cite{Vries2013} in a way that can differ from EDMs\cite{Liu2012}.


The measurement procedure for a nuclear MQM (nMQM) is similar to the electron EDM (eEDM), where the strategy is to take advantage of the internal molecular electromagnetic field to further enhance the effect of the nMQM. Thus, one would naturally look for diatomic molecules already employed in eEDM experiments that contain heavy quadrupole-deformed nuclei. 
An alternative is to turn towards the analogous triatomic molecules that were recently identified as even more promising candidates for the search for $\mathcal{P,T}$-odd interactions \cite{Kozyryev2017,denis2019,Isaev2017RaOH}. Due to their favourable vibrational structure, these molecules possess internal comagnetometer states and they can be fully polarised in comparatively low electric fields.

In the present work, we aim to determine the nMQM interaction constant of the laser-coolable di- and triatomic molecules BaF, YbF, BaOH, and YbOH with a highly accurate relativistic approach to molecular structure. In addition, we provide an estimate of the errors of our results. The choice of ytterbium is motivated by the promising deformed shape of its nucleus, while barium compounds were added to provide an additional benchmark with previous works. Previous results for our systems of interest were obtained in a semi-empirical way by evaluating the nMQM interaction constants from hyperfine interaction constants \cite{Kozlov1995}, or by employing various Hartree-Fock approaches\cite{Quiney1998,Titov1996} for BaF and YbF. A recent result on YbOH was obtained by making use of the relativistic coupled cluster approach\cite{Maison2019}. We shall compare the outcomes of our study with these works.

\section{Methodology}
The nuclear magnetic quadrupole moment interaction is described by the Hamiltonian\cite{Ginges2004,Skripnikov2017}
 \begin{align}
 \label{hmqm}
     H_\text{MQM}=-\frac{M}{2I(2I-1)}T_{ik}\cdot\frac{3}{2}\frac{[\boldsymbol{\alpha}\times\boldsymbol{r}]_i}{r^5}r_k,
 \end{align}
 where $\boldsymbol{\alpha}$ are the $4\times4$ Dirac matrices, $\boldsymbol{r}$ is the position of the electron with respect to the considered nucleus, $M$ is the nMQM, and we introduced the second-rank tensor $\boldsymbol{\hat{T}}$ with components $T_{ik}=I_iI_k+I_kI_i-\frac{2}{3}I(I+1)\delta_{ik}$, with $I$ the spin of the nucleus.

 In the subspace of the $\pm \Omega$ states where $\Omega$ is the projection of the total electronic angular momentum along the molecular axis, \cref{hmqm} can be reduced to an effective Hamiltonian\cite{Sushkov1984}
 \begin{align}
     H_\text{MQM}^\text{eff}=-\frac{W_MM}{2I(2I-1)}\boldsymbol{S}'\boldsymbol{\hat{T}}\boldsymbol{n}  ,
 \end{align}
 with $\boldsymbol{n}$ the unit vector along the internuclear axis and $\boldsymbol{S'}$ the effective electron spin \cite{Kozlov1995}, which in the case of the molecular axis coinciding with the z-axis obeys $\boldsymbol{S}'_z| \Omega\rangle = \Omega|\Omega\rangle $. The molecular energy shift then reads
\begin{align}
 \Delta E_\text{MQM}
&=-\frac{1}{3}W_MM\Omega, 
\end{align}
where the molecular interaction constant $W_M$  is given by an expectation value
\begin{equation}
 W_M=\frac{3}{2\Omega}\left\langle \sum_{j=1}^n \left( \frac{\boldsymbol{\alpha}_j\times\mathbf{r}_{jA}}{r_{jA}^5} \right)_{k=3} \left(r_{jA}\right)_{k=3}  \right\rangle_{\psi_\Omega}.
 \end{equation}
 Here, rather than calculate the expectation value explicitly, we make the choice to use the finite-field approach to obtain this property\cite{Coh65,PopMcIOst68,Mon77}. 
 In this framework, the nMQM Hamiltonian is added as a perturbation to the relativistic Dirac-Coulomb Hamiltonian,
  \begin{equation}
 H^{(0)}=\sum_i^n\left[c\boldsymbol{\alpha}_i\cdot\textbf{p}_i+\beta_i c^2+V_{\text{nuc}}(\textbf{r}_i)\right]
 +\sum_{i<j}\frac{1}{r_{ij}}.
 \end{equation}
 We replace $M$ by a parameter $\lambda$ in the nMQM Hamiltonian to make explicit that we treat this interaction as a perturbation. 
 Provided that the values of the $\lambda$ are small enough to ensure a linear behaviour of the energy, the nMQM interaction constant $W_M$ can then be obtained numerically, according to the Hellmann-Feynman theorem, from the first derivative of the energy with respect to $\lambda$. Namely, in the limit of the exact wave function $W_M$ can be related to the expectation value
 \begin{equation}
\label{wm_ff}
W_{M}=\bra{\Psi}  H^{'}_\text{MQM} \ket{\Psi} \simeq \left.\frac{dE(\lambda)}{d \lambda}\right|_{\lambda=0}.
\end{equation}
Here, the perturbative nMQM Hamiltonian reads
\begin{equation}
    H^{'}_\text{MQM}=\frac{3}{2\Omega} \sum_{j=1}^n \left( \frac{\boldsymbol{\alpha}_j\times\mathbf{r}_{jA}}{r_{jA}^5} \right)_{k=3} \left(r_{jA}\right)_{k=3} .
\end{equation}
It can be rewritten in terms of the electric-field gradients, which is simpler for implementation,  by making use of the equalities
\begin{equation}
 \left(\frac{\boldsymbol{\alpha}\times\mathbf{r}}{r^5}\right)_{3}r_3=\alpha_1\frac{r_2 r_3}{r^5}-\alpha_2\frac{r_1 r_3}{r^5},
 \end{equation}
 
 and
\begin{equation}
 \frac{r_i r_j}{r^5} =\frac{-1}{3q}\frac{\partial}{\partial r_i}E_j(\boldsymbol{r}),
\end{equation}
where $E_j(\boldsymbol{r})$ is the electric field at the position of the electron $j$ with charge $q$. Thus, the perturbative Hamiltonian as implemented in our code reads
\begin{equation}
 H^{'}_\text{MQM}=\frac{3}{2\Omega}\frac{1}{3}\left( \alpha_1\partial_2 E_3 -\alpha_2\partial_1E_3 \right).
\end{equation}

We also perform calculations of the parallel magnetic  hyperfine  structure (HFS) constant, defined as the expectation value of the projection of the magnetic vector potential along the molecular axis,
\begin{equation}
    A_{||}=\frac{\mu_z}{I\Omega}\left\langle\sum_{i=1}^{n}\left(\frac{\boldsymbol{\alpha}_i\times\mathbf{r}_{iA}}{r_{iA}^3}\right)_z\right\rangle_\psi,
\end{equation}
where $\mu_z$ is the magnetic moment of the considered atom.
The corresponding Hamiltonian is added as a perturbation to the unperturbed Dirac-Coulomb Hamiltonian $H^{(0)}$ and $A_{||}$ is obtained through the finite-field procedure described above (see \cref{wm_ff}).  

Finally, another quantity of interest for future experimental developments is the nuclear quadrupole coupling constant (NQCC), which reads 
\begin{equation}
 \text{NQCC}=eq_0Q,
\end{equation}
where $Q$ is the nuclear quadrupole moment, $q_0$ is obtained by evaluating the electric-field gradient EFG along the internuclear axis at the position of the nucleus $\boldsymbol{r}_0$,
\begin{equation} \left. 
q_{zz}(r_0)=\frac{\partial^2V(\boldsymbol{r})}{\partial z^2} \right|_{r_0}.
\end{equation}
The electronic part is obtained by employing the finite-field method where the operator 
\begin{equation}
q^{elec}_{zz}(r_0)=\sum^{n}_{i=1}\frac{3(z_i-z_0)^2-|\boldsymbol{r}_i-\boldsymbol{r}_0|^2}{|\boldsymbol{r}_i-\boldsymbol{r}_0|^5}.
\end{equation}
is added as a perturbation to the Dirac Hamiltonian. 

\section{Computational details}
The present work was performed using the DIRAC17 program package \cite{DIRAC17}. For all the molecules, we employed experimental equilibrium geometries. The four systems are linear in their $X^2\Sigma$ ground-state and display the following atomic distances: d$_\text{Ba-F}=2.162$\AA{} for BaF  \cite{Knight1971}, d$_\text{Ba-O}=2.201$\AA{}, d$_\text{O-H}=0.923$\AA{} for BaOH 
\cite{KinseyNielsen1986},  d$_\text{Yb-F}=2.016$\AA{} for YbF \cite{Barrow1975} and  
d$_\text{Yb-O}=2.0369$\AA{}, d$_\text{O-H}=0.9511$\AA{} for YbOH \cite{Brutti2005}.

In our calculations of the $W_M$ and the $A_{||}$ parameters, the finite-field approach, as described above, was employed. 
To this end, the same calculation was repeated three times, with $\lambda$  factors of $-10^{-7}$, $0$, and $1.10^{-7}$, except in the case of the HFS constant in BaF and BaOH, where $-10^{-6}$, $0$, and $1.10^{-6}$ $\lambda$ were applied. These field strengths were chosen to ensure linear response of the total energy to the perturbation, so that \cref{wm_ff} applies. In view of such small field strengths, the convergence criterion of the coupled cluster amplitudes was set to $10^{-12}$ a.u.. 
The nMQM interaction constant $W_M$ was then obtained as the derivative of the energy with respect to the field strength from linear fitting of the three energy points. The hyperfine structure constant was derived in a similar manner.

The electron correlation was treated via relativistic coupled cluster approaches. We used both the standard single-reference coupled cluster with single, double, and perturbative triple excitations (CCSD(T)) \cite{VisLeeDya96}  and the multireference Fock-space coupled cluster (FSCC) \cite{Visscher2001}. The uncorrelated Dirac-Hartree-Fock (DHF) and the 
second-order many-body perturbation theory  (M\o{}ller-Plesset theory, MP2) \cite{MolPle34} values are given for the purpose of comparison.
In the correlated calculations, if not stated otherwise, all the electrons were correlated and the virtual space cutoff was set to 2000 a.u.

 Dyall's relativistic uncontracted basis sets of varying quality (double to quadruple-zeta) \cite{Dyall2009,GomDyaVis10,Dyall2016} were employed in the calculations. We also test the effect of augmentation of these standard basis sets by further diffuse (low exponent) and tight (high exponent) basis functions. The tight functions in particular may prove to have a significant effect on the calculated  $W_M$ constants, as they improve the description of the wave function in the nuclear region.  

\section{Results and discussion}
In the following, we investigate the effect of various computational parameters on the calculated nMQM interaction constants (Sections C.1-C.4), in order to determine the best computational model for each system and to estimate the uncertainty of the values obtained with this model (Section C.5). The final results with their assigned error bars are then compared to the previous investigations (Section D). 


\subsection{Basis set size}
\begin{figure}
    \centering
    \includegraphics[width=\columnwidth]{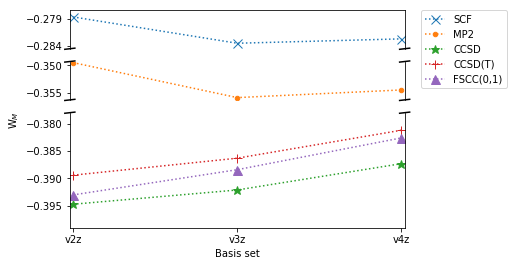}
    \caption{Influence of the size of the basis set quality on the calculated $W_M$ values of BaF. Results are given in $[\frac{10^{33}{\rm Hz}}{{\rm e.cm}^2}]$.}
    \label{BaFbasis}
\end{figure}

\begin{figure}[h]
    \centering
    \includegraphics[width=\columnwidth]{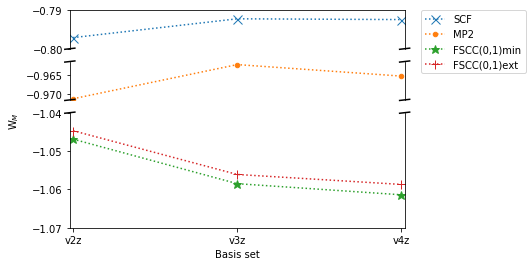}
    \caption{Influence of the size of the basis quality  set on the calculated W$_M$ values of YbF. Results are given in $[\frac{10^{33}{\rm Hz}}{{\rm e.cm}^2}]$.}
    \label{YbFbasis}
\end{figure}

\cref{BaFbasis} and \cref{YbFbasis}  display the behaviour of the $W_M$ constants in BaF and YbF, calculated on various levels of theory with increasing basis set size.  Both molecules display a non-monotonic behaviour of the SCF results, which decrease from v2z to v3z and slightly increase towards v4z for BaF, and show the opposite behaviour for YbF. The trends in the triatomic molecules are identical to those in their diatomic homologues, and hence are not presented in the Figures.
\Cref{BaFbasis} also highlights the anomalous trend in the post-Hartree Fock correlated results for BaF (\textit{i.e.}, CCSD and CCSD(T)), which is highly inconsistent with the usually observed convergence of molecular properties with the basis set size. For Yb (\cref{YbFbasis}), however, we see the expected behaviour.   
\cref{tabbasisset} contains the CCSD(T) results for both BaF and BaOH. For BaF, when going from the v3z to the v4z basis set, the $W_M$ undergoes a $1.3\%$ increase,  significantly larger than the $0.8\%$ variation when switching from the v2z to the v3z basis. Similar trend is observed for BaOH. Such an anomalous trend was already seen for another $\mathcal{P,T}$-odd property, the eEDM enhancement factor ($E_\text{eff}$) in Ref. \cite{denis2019}, which also undergoes an increase and then a decrease upon enlarging the basis set.
A possible explanation for the lack of saturation with respect to the basis set could be in that these basis set were not optimised for calculations of properties on the coupled cluster level. 
As a consequence, it would be unwise to attempt extrapolation to the complete basis set limit (CBSL) for the barium compounds. 
\begin{table}[h]
   \caption{Calculated $W_M$ (\unitWm) with increasing quality basis sets. The results were obtained with a $2000$ a.u. virtual cutoff, all electrons correlated, and the CCSD(T) method for the Ba compounds and the FSCC(0,1) Ext method for YbF and YbOH.}
    \label{tabbasisset}
    \centering
    \begin{tabular*}{\columnwidth}{l@{\extracolsep{\fill}}cccc}
    \hline 
    \hline              
    Basis set& W$_M$(BaF)&W$_M$(BaOH)& W$_M$(YbF)&W$_M$(YbOH)\\
    \hline
   v2z    & $-0.3894$ & $-0.3955$ & $-1.0447$ & $-1.0448$\\    
    v3z    & $-0.3863$ & $-0.3934$ & $-1.0561$& $-1.0629$\\    
    v4z    & $-0.3812$ & $-0.3885$ & $-1.0587$ &$-1.0686$\\
    CBSL    & & & $-1.0606$ & $-1.0728$ \\
    \hline
    \hline
    \end{tabular*}
\end{table}

In contrast, the FSCC results for YbF and YbOH (\cref{tabbasisset}) behave as expected, namely, they exhibit a converging trend with respect to the enlargement of the basis set. We can thus extrapolate these results to the complete basis set limit by using the inverse cubic extrapolation scheme for the coupled cluster energies\cite{Helgaker1997}. The complete basis set limit values are very close to the v4z results, which indicates a good convergence of $W_M$ with respect to the size of the basis set. This convergence is slightly better in the case of YbF than in YbOH, with CBSL and v4z values at a variance of $0.2\%$ and $0.4\%$ in magnitude, respectively. 

We now check the saturation of the v4z basis sets by testing the effect of adding extra basis functions, starting with diffuse functions. We employ the s-augmented Dyall-v4z basis set, which is generated by the Dirac program by supplementing the dyall-v4z basis with a single diffuse function for each symmetry. As displayed in \cref{tabv4z}, this results in a minuscule change below $0.05\%$ for the  $W_M$ constants of YbF and YbOH and a larger but still small variation of $0.4\%$ in magnitude for those of BaF and BaOH. This study indicates the comprehensiveness of the Dyall valence basis set in regard to small exponent basis functions. 

We proceed to explore the effect of the inclusion of extra tight functions. These functions are expected to play a more important role than the diffuse ones, since the parity-violating property we aim to evaluate is known to be highly dependent on the electron spin density in the vicinity of the heavy nucleus and thus rather sensitive to the description of the electronic wavefunction in the nuclear region. To check the effect of the tight functions, we used Dyalls core-valence (cv4z) basis sets, which include six Ba (3f, 2g, 1h), three O (2d, 1f), and three F (2d, 1f) additional tight functions.
Dyall's valence basis sets (vNz) for lanthanides, and thus for Yb, already include the core-valence functions and are \textit{de facto} identical to the cvNz basis sets (these basis sets were optimised to provide a good description of atoms where the 4f shell is close in energy to the valence orbitals).

The results in \cref{tabv4z} show a significant $1.3\%$ decrease of W$_M$ in BaF and BaOH, confirming the need for a high quality description of the region around the nucleus. 
In view of the importance of the effect, we further explore the saturation of the cv4z basis sets by using the Dyall all-electron quadrupole-zeta basis sets, denoted ae4z, which includes nine additional Ba (5f, 3g and 1h) and five additional Yb (1f, 3g, 1h) tight functions. The effect is there minuscule, below $0.03\%$ for all the systems, confirming the saturation of the cv4z basis set.

We can conclude from the basis set analysis that the optimal choices of the basis sets are the cv4z  for the Ba compounds  and the complete basis set limit extrapolation of the vNz basis sets for the Yb compounds. 


\begin{table}[h]
   \caption{Calculated $W_M$ (\unitWm) with various augmentations of the v4z basis sets. The results were obtained with a $2000$ a.u. virtual cutoff, all electrons correlated, and the CCSD(T) method for the Ba compounds and the FSCC(0,1) Ext approach for YbF and YbOH.
   }
    \label{tabv4z}
    \centering
    \begin{tabular*}{\columnwidth}{l@{\extracolsep{\fill}}cccc}
    \hline 
    \hline              
    Basis set& W$_M$(BaF)&W$_M$(BaOH)& W$_M$(YbF)&W$_M$(YbOH)\\
    \hline
    v4z     & $-0.3812$  &  $-0.3885$ & $-1.0614$ &$-1.0712$ \\
    s-aug-v4z & $-0.3795$ & $-0.3868$ & $-1.0609$ &$-1.0707$\\
    cv4z    & $-0.3862$& $-0.3933$&  &\\
    ae4z    & $-0.3861$& $-0.3932$& $-1.0615$ & $-1.0714$\\
  
    \hline
    \hline
    \end{tabular*}
\end{table}

\subsection{Treatment of relativity}
In order to account for the electron-electron interaction beyond the Coulomb approximation, one should consider the Breit term that corrects the 2-electron part of the Dirac Coulomb Hamiltonian up to order $(Z\alpha)^2$,
\begin{align}
    g^{{\rm Breit}}\left(1,2\right)&=-\frac{c{\alpha}_{1}\cdot c{\alpha}_{2}}{2c^{2}r_{12}}-\frac{\left(c{\alpha}_{1}\cdot\boldsymbol{r}_{12}\right)\left(c{\alpha}_{2}\cdot\boldsymbol{r}_{12}\right)}{2c^{2}r_{12}^{3}}.
\end{align}
 The Breit operator can be split into two contributions respectively called Gaunt or magnetic term and gauge term, {\it viz.}
\begin{align} g^{{\rm Breit}}\left(1,2\right)&=-\frac{c{\alpha}_{1}\cdot c{\alpha}_{2}}{2c^{2}r_{12}}-\frac{\left(c{\alpha}_{1}\cdot\boldsymbol{r}_{12}\right)\left(c{\alpha}_{2}\cdot\boldsymbol{r}_{12}\right)}{2c^{2}r_{12}^{3}}\nonumber
\\&=-\frac{c{\alpha}_{1}\cdot c{\alpha}_{2}}{c^{2}r_{12}}-\frac{\left(c{\alpha}_{1}\cdot{\nabla}_{1}\right)\left(c{\alpha}_{2}\cdot{\nabla}_{2}\right)r_{12}}{2c^{2}}\nonumber
\\&= g^{{\rm Gaunt}}\left(1,2\right)+g^{{\rm gauge}}\left(1,2\right).\end{align}
The DIRAC code\cite{DIRAC17} restricts the 2-electron correction to the Gaunt term at the DHF level since it was shown to be dominant in atomic calculations \cite{Grant1970,visscher1993} and is more easily implementable than the complete operator.
This term completes the Coulomb term by adding the treatment of spin-other orbit interaction to the already included spin-same orbit interaction.
Including the Gaunt term slightly lowers $W_M$ by $0.3\%$ for BaF and BaOH and $0.5\%$ for YbF and YbOH.

The effect of the Gaunt contribution on the nuclear magnetic quadrupole moment interaction constants is much smaller than that on the effective electric fields in the same systems, where it reached $1.7\%$ and $1.5\%$ for the baryum and ytterbium compounds, respectively\cite{denis2019}.

\subsection{Virtual space cutoff}
The virtual space cutoff is usually set at $30$ a.u. for the calculation of standard (valence) properties. In our work, however,  this cutoff is raised to $2000$ a.u., for it is known that high lying virtual orbitals are critical for the correlation of the core electrons in evaluation of $\mathcal{P,T}$-odd properties. We also investigate the effect of further increasing the virtual space cutoff at the triple-zeta level.   
In case of YbF and YbOH, the inclusion of 24 extra spinors to reach $6000$ a.u. brings a small $0.3\%$ decrease of the  $W_M$ value, and further increasing the cutoff until $10000$ a.u., \textit{i.e.}  adding 18 more spinors,  alters the result by an extra $0.17\%$.
BaF and BaOH, we observe a similar trend: going from a $2000$ a.u. to a $6000$ a.u. cutoff lowers the  $W_M$ value by $0.23\%$ and going further to $10000$ a.u. brings another $0.20\%$ decrease. Further extension of the virtual space (to $20000$ a.u.) has negligible effect on the calculated $W_M$.

\subsection{Treatment of electron correlation}

\begin{table}[htb]    
\caption{Calculated values of $W_M$ in BaF and BaOH obtained with Dyall-cv4z basis set, all electrons correlated, a 2000 a.u. virtual space cutoff and different correlation methods.}
    \centering
    \begin{tabular*}{\columnwidth}{@{\extracolsep{\fill}}lcc}
    \hline \hline
    Method & W$_M$(BaF)&W$_M$(BaOH)\\
    \hline
    DHF     & $-0.2829$ & $-0.2848$\\
    MP2     & $-0.3547$ & $-0.3571$\\
    CCSD   & $-0.3924$ & $-0.3995$\\
    CCSD(T)   & $-0.3862$ & $-0.3933$ \\
    CCSD+T   & $-0.3856$ & $-0.3929$\\
    CCSD-T   & $-0.3869$ & $-0.3938$\\
    FSCC(0,1) Min   &$-0.3874$ & $-0.3936$\\
    FSCC(0,1) Ext   & $-0.3861$ & $-0.3921$ \\
    FSCC(1,0)   & $-0.3956$ & $-0.4045$ \\    
    \hline\hline
    \end{tabular*}

    \label{tabcorrBa}
\end{table}
In order to highlight the crucial character of the treatment of electron correlation in the evaluation of $\mathcal{P,T}$-odd properties, we carry out calculations at various levels of correlation. Results  obtained with the optimal basis sets defined above are displayed in \cref{tabcorrBa,tabcorrYb}. In all the systems considered in this work, the omission of correlation, as done in DHF calculations, entails a $26$ to $29\%$ error compared to the all-electron correlation method CCSD, while use of the MP2 method drives the error down to $10\%$. 
The treatment of triple excitations for the single reference coupled cluster approach is implemented in a perturbative way in the DIRAC program, following three various schemes, namely CCSD(T), CCSD+T and CCSD-T respectively. The standard scheme (CCSD(T), \cite{RagTruPop89}) includes all fourth-order terms and part of the fifth-order terms while CCSD-T \cite{DeeKno94} includes further fifth-order terms and CCSD+T includes fourth-order terms only \cite{UrbNogBar85}. The three schemes are in excellent agreement (within $0.3\%$ of each other) for BaF and BaOH, increasing the total $W_M$ values by $1.6\%$, compared to the CCSD results.

The multireference character of the ytterbium compounds, leads to highly unreliable results of the triple excitation schemes as well as to irregular behaviour of single reference CCSD with respect to basis set. This echoes the findings in previous studies of the $E_\text{eff}$ in YbF and YbOH \cite{Abe2014,denis2019} .
\begin{table}[h] 
\caption{Calculated values of $W_M$ in YbF and YbOH obtained with Dyall-v4z basis set, all electrons correlated, a 2000 a.u. virtual space cutoff and different correlation methods.}
    \centering
    \begin{tabular*}{\columnwidth}{@{\extracolsep{\fill}}lcc}
    \hline \hline
    Method & W$_M$(YbF)&W$_M$(YbOH)\\
    \hline
    DHF     & $-0.7924$ & $-0.7986$\\
    MP2     & $-0.9653$ & $-0.9717$\\
    CCSD   & $-1.0732$ & $-1.0867$\\
    FSCC(0,1) Min   & $-1.0614$ & $-1.0712$ \\
    FSCC(0,1) Ext   & $-1.0587$ & $-1.0686$\\
    FSCC(1,0)   & $-1.0965$ & $-1.1159$ \\    
    \hline\hline
    \end{tabular*}
    \label{tabcorrYb}
\end{table}

As a test of the reliability and consistency of the coupled cluster correlation method for the calculation of $W_M$ in BaF and BaOH, and because CCSD(T) approach is not applicable to YbF and YbOH, we employ an alternative coupled cluster method, namely the Fock Space coupled cluster, FSCC. 
In the case of $^2\Sigma_{\frac12}$ molecules that have an unpaired $\sigma$ electron, two computational schemes are appropriate. The first one is denoted FSCC(0,1); here, the calculation starts from the BaF$^+$, BaOH$^+$, YbF$^+$, or YbOH$^+$ cations and then an electron is added in the coupled cluster procedure. The set of orbitals in which the electron is permitted to enter, the model space, has to be defined. In the present work, two model spaces were employed, the minimal one, denoted Min, which only includes the lowest $\sigma$ orbital, and the extended one, denoted  Ext, which includes the twelve lowest spinors, (\textit{i.e.}, 2$\sigma$, 2$\pi$, 2$\delta$ orbitals).  
The second computational scheme is denoted FSCC(1,0) and starts from the anions BaF$^-$, BaOH$^-$, YbF$^-$, or YbOH$^-$, from which an electron is removed within the calculation.
Results shown in \cref{tabcorrBa} highlight the remarkable performance of the FSCC(0,1) scheme that returns values very close (within $0.3\%$) to those obtained with CCSD(T), despite the absence of explicit treatment of triple excitations. This feature was previously observed \cite{EliBorKal17}, and notably in the studies of parity violating properties \cite{HaoIliEli18,denis2019}. Furthermore, the consistency of the Fock space coupled cluster approach is attested by the excellent agreement of the two model spaces in the $0.2\%-0.4\%$ range both in barium and ytterbium compounds.
On the other hand, the FSCC(1,0) scheme does not perform as well as FSCC(0,1), and these results are at a discrepancy larger than $2.5\%$ with respect to the CCSD(T) for BaF and BaOH.
This poor behaviour was also observed in the study of the electron EDM enhancement factor in the same systems \cite{denis2019} and attributed to the unsuitability of the employed basis sets for the description of the negatively charged ions. 

Consequently, we will use the CCSD(T) results for the final $W_M$  values of BaF and BaOH, and the FSCC(0,1) scheme with the extended model space for YbF and YbOH. 

\subsection{Recommended values and error estimation}
In the analysis above we found that the optimal model for the calculation of $W_M$ in BaF and BaOH is the CCSD(T) method with a 2000 a.u. virtual space cutoff, all electron correlated, and the core-valence quadruple-zeta basis sets. As for YbF and YbOH, the optimized scheme
proved to be the FSCC(0,1) Ext method with a 2000 a.u. virtual space cutoff, all electron correlated, and the valence basis sets extrapolated to the complete basis set limit. \cref{tabfinal} contains the results  obtained with these models and corrected for the Gaunt contribution.
It is interesting to note that the ratio of the constants of YbF and BaF is about 2.7 versus 3.6 for $E_{\rm eff}$. The expected scaling is~\cite{Sushkov1984,Flambaum2014} $\sim Z^2$ for MQM versus $Z^3$ for eEDM, so one expects the ratio to be roughly $\sim Z^{2/3}$, fairly close to what is seen.

\begin{table}[hbt]
 \caption{Summary of the most significant error sources in the calculated $W_M$ constants in BaF, BaOH, YbF, and YbOH, given in \unitWm.
}
 \label{taberror}
 \begin{tabular*}{\columnwidth}{l@{\extracolsep{\fill}}lcccc}
 \hline\hline
\multicolumn{2}{l}{Error source}	    	& BaF & BaOH & YbF			&	YbOH \\
\hline
\multicolumn{2}{l}{Basis quality}    &		$0.0051$		 & $0.0049$	& $ 0.0019$ & $ 0.0042$	\\
\multicolumn{2}{l}{Basis augmentations}	&			 	& &	&	\\
&Tight functions&$0.0001$	& $0.0001$ & $0.0001$ & $0.0002$ \\
&Diffuse functions	 &$0.0017$& $0.0017$ &$0.0005$& $ 0.0005$ \\
\multicolumn{2}{l}{Correlation} & & & &\\
&Virtual space cut-off& $0.0034$&$0.0034$ & $0.0096$& $0.0096$\\
&Residual triples and &	&	&  & \\
&higher excitations	    & $0.0026$		&  $0.0018$	& $0.0378$ &  $0.0473$	\\
\multicolumn{2}{l}{Relativity} & $0.0013$ & $0.0013$ & $0.0053$ & $0.0053$  \\ 

\cline{3-6}
\multicolumn{2}{l}{\textbf{Total~~~}}	    & $0.0070$		&	$0.0066$	&  $0.0394$   &			$0.0487$	
\\
\hline
\hline
 \end{tabular*}
\end{table}
The nMQM interaction constant $W_M$ can not be measured, and thus an important part of this work is to assign reliable uncertainties on our calculated values. We base these uncertainties on the extensive and comprehensive scrutiny of the effect of the various computational parameters that we performed in the previous sections. 

The main remaining sources of error in our calculations are the incompleteness of the basis set, the neglect of higher order relativistic effects (beyond the Gaunt contribution) and the neglect of higher excitations and the virtual space cutoff in the correlation treatment. Below, we address these sources of error separately.

The basis sets used for the final values are likely saturate; nonetheless we estimate the possible effect of further increasing the quality the basis set by taking the difference between the quadruple-zeta and the triple-zeta basis set for BaF and BaOH, and between the complete basis set extrapolation and the quadruple-zeta in the case of YbF and YbOH.
The uncertainty on the saturation of the basis set with respect to the tight and diffuse functions is assumed to be smaller than the correction brought by the augmented basis set, (\textit{i.e.} all electron and s-augmented, respectively) to the optimal basis set. 
The corresponding values are compiled in \cref{taberror}.

The $2000$ a.u. virtual space cutoff set in most calculations of this work, albeit much higher than in standard relativistic calculations, is finite and thus remains an approximation. As a conservative estimate of the corresponding uncertainty we take double the difference between the values of $W_M$ obtained with a cutoff of $10000$ and $2000$ a.u. This estimate is justified by the assumption that the correction due to a further enlargement of the virtual space should not be larger than the $2000-10000$ a.u. difference. Hence, the uncertainty due to the truncation of the virtual space is around $0.9\%$, depending on the system considered.

In the barium compounds, it was possible to treat the triple excitations in a perturbative way. To evaluate the uncertainty due to incomplete treatment of triple and higher excitations, we use the spread in the values obtained with different schemes of inclusion of the triple excitations, namely, we take twice the difference between CCSD+T and CCSD-T values. For YbF and YbOH, no direct evaluation of the triple excitations was feasible because the perturbative treatment proved to be unreliable. In this case we assign the uncertainty by comparing the $W_M$ constants obtained using the two Fock space coupled cluster schemes.  
This procedure leads for YbF and YbOH to the respective uncertainty of  $3.5\%$ and $4\%$; these conservative figures are consistent with the complex electronic structure of the
ytterbium atom.

The final contribution to the uncertainty originates in the neglected QED effects. We assume that the Gaunt term accounts for most of the Breit correction and hence the missing part should not alter the results by more than the Gaunt term itself. Therefore, uncertainty due to the missing effects will be estimated by the magnitude of the Gaunt contribution.  

\begin{figure}[htb]
    \centering
    \includegraphics[width=0.49\textwidth]{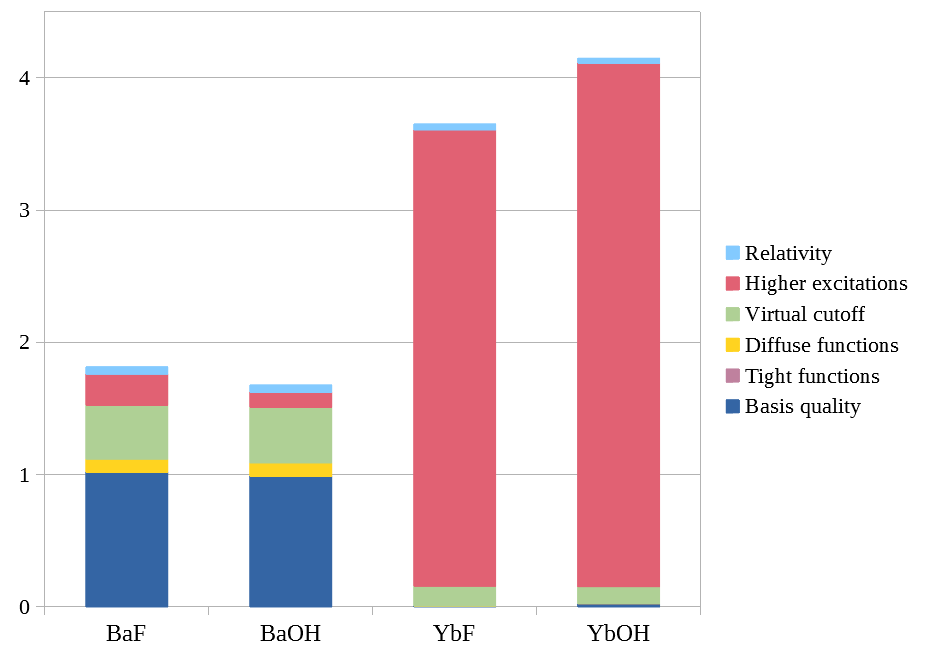}
    \caption{Contribution of the different error sources to the total error bar in each system given in percentage of the final values.}
    \label{figerror}
\end{figure}
We assume the various sources of error to be independent and obtain the total uncertainty as the square root of the sum of the constituent uncertainties. 
\cref{figerror} illustrates the contribution of the different parameters to the total error. For the ytterbium compounds the total uncertainty of about $4\%$ is dominated by the lack of treatment of triple and higher excitations.
On the other hand, for BaF and BaOH, the largest source of error is due to  the impossibility of performing a complete basis set extrapolation. Nonetheless, the total uncertainty for the Ba compounds is lower, at about $2\%$. The present error bars are of similar magnitude to those determined in our previous work\cite{denis2019} on the effective electric field in BaOH and YbOH. The final recommended values along with their absolute error bars are compiled in \cref{tabfinal}.

\begin{table}[h]  
\caption{Final recommended values of  $W_M$ along with the estimated error bars given in \unitWm .}
    \label{tabfinal}
    \centering
    \begin{tabular*}{\columnwidth}{l@{\extracolsep{\fill}}lcl}
    \hline\hline
         System& Model & W$_M$&\\
         \hline
         BaF & cv4z CCSD(T)+gaunt  & $-0.385 \pm 0.007$ & \\
         BaOH & cv4z CCSD(T)+gaunt  & $-0.392 \pm 0.007$ & \\
         YbF  & CBSL FSCC(0,1)Ext+gaunt  & $-1.055\pm 0.039$&\\
         YbOH  & CBSL FSCC(0,1)Ext+gaunt  & $-1.067\pm 0.049$ & \\

         \hline\hline
    \end{tabular*}
 
\end{table}
We also calculated the parallel magnetic HFS constant by employing the computational models optimized for the calculation of the $W_M$ constants. 
 The aim of this calculation is twofold: besides providing a useful spectoscopic parameter, we can also use it as a test of our method. 
Indeed, the parallel magnetic hyperfine interaction constant exhibits similar features as the $\mathcal{P,T}$-odd constants in that they are both sensitive to the electron spin density around the nucleus. However, unlike $W_M$ or ${\cal E}_\text{eff}$, A$_{||}$ can be measured and thus the calculated values can be compared to experimental data.  
\cref{tabhyp} contains the calculated magnetic parallel hyperfine interaction constants as well as the experimental values when available and the relative discrepancy between them. For ytterbium, we employed both $^{171}$Yb and $^{173}$Yb isotopes because there are experimental data for the first one and the second is the isotope of interest for the search for the nMQM. To extract the hyperfine constants, the following magnetic moments
\cite{stone2014} were employed
$\mu(^{137}\text{Ba})=0.93737
\mu_{N}$, $\mu(^{171}\text{Yb})=0.49367\mu_{N}$, $\mu(^{173}\text{Yb})=-0.67989
\mu_{N}$.

\begin{table}[h]  
\caption{The $A_{||}$ constants calculated using the model optimized for $W_M$ and comparison with the available experimental data.}
    \label{tabhyp}
    \centering
    \begin{tabular*}{\columnwidth}{l@{\extracolsep{\fill}}rrr}
    \hline\hline
         System& Calculated A$_{||}$& Experimental A$_{||}$ & $\Delta$($\%$) \\
         \hline
         $^{137}$BaOH &  $2194.6$ &  $2200.2$\cite{Anderson1993}& $0.3\%$\\
         $^{171}$YbF  & $7579.0$ & $7505.9$\cite{Steimle2007} & $1\%$\\
         $^{173}$YbF  & $-2087.6$ & $-2085.1$\cite{Wang2019} & $0.1\%$\\
         $^{171}$YbOH & {$7174.9$} &  & \\
         $^{173}$YbOH & {$-1976.3$} & &\\
         \hline\hline
    \end{tabular*}

\end{table}

The results display an excellent agreement with the experimental values, with a deviation below $1\%$, well within the uncertainty evaluated for the W$_M$ final values. Our confidence in the accuracy of our method is thus reinforced.  

Finally, the calculated effective field gradients of the triatomic molecules are displayed in \cref{tabEFG} along with the ensuing NQCC. The nuclear moments $Q$ employed are\cite{Wendt1984,Zehnder1975}: $Q$(Ba) = 245 mb and $Q$(Yb) = 2800 mb.    
\begin{table}[h]
    \centering
    \begin{tabular*}{\columnwidth}{l@{\extracolsep{\fill}}cr}
    \hline \hline
    System   & EFG [a.u.]  & NQCC [MHz] \\
    \hline 
    $^{137}$BaOH     &  \com{$-2.776$}& \com{$-159$}\\
    $^{173}$YbOH    & \com{$-5.367$} &  \com{$-3551$} \\
    \hline \hline
    \end{tabular*}
    \caption{Calculated effective field gradients and corresponding NQCC of BaOH and YbOH obtained with models optimized for $W_M$. }
    \label{tabEFG}
\end{table}

\subsection{Comparison with previous work}
There are very few available data for the nMQM interaction constant for $^2\Sigma_{\frac12}$ molecules and, with the exception of the recently published work on $W_M$ in YbOH, none obtained with an accurate computational method. 
The most striking point of \cref{tabcomp} is the significant discrepancy of our values compared to those obtained by Kozlov\cite{Kozlov1995} through a semi-empirical procedure. This method performed reasonably well for other $\mathcal{P,T}$-odd interaction constants, namely the effective electric field $E_\text{eff}$ \cite{Kozlov1995,denis2019}. Combined with the fact that the ratio of the semiempirical $W_M$ constants for the Yb and the Ba compounds is very similar to that obtained here, this leads us to suspect a missing factor of 2 in the definition used in Ref. \cite{Kozlov1995}.
Our results for YbF are in reasonable agreement with the earlier Dirac-Hartree-Fock\cite{Quiney1998} and RASSCF\cite{Titov1996} values.
Finally, an almost perfect agreement is achieved when compared to the most recent study of W$_M$ in YbOH\cite{Maison2019}, which was to be expected, since the procedure and method are very similar. Still, in their work the virtual cutoff is lower than in our case and all the electrons are not treated on an equal footing. The core electrons (Yb \textit{1s} to \textit{2p}) are frozen and their contribution evaluated by comparison within a small basis set of a double-zeta quality.

\begin{table}[h]
    \centering 
    \caption{Final recommended $W_M$ values and comparison with previous work, given in $[\frac{10^{33}{\rm Hz}}{{\rm e.cm}^2}]$.}
    \label{tabcomp}
    \begin{tabular*}{\columnwidth}{l@{\extracolsep{\fill}}llr} 
    \hline \hline
    \multicolumn{2}{l}{System} &Method & W$_M$\\    
    \hline
 \multicolumn{2}{l}{BaF}  & & \\
      & \textbf{This work}  & \textbf{CCSD(T)+gaunt} & $\boldsymbol{-0.385}$\\
      &\cite{Kozlov1995}Kozlov, 1995 & Semi-empirical from HFS\cite{Knight1971}& $-0.83$ \\
       &\cite{Kozlov1995}Kozlov, 1995 & Semi-empirical from HFS\cite{Ryzlewicz1982}& $-0.98$ \\     
\multicolumn{2}{l}{BaOH}  & & \\
     & \textbf{This work}  & \textbf{CCSD(T)+gaunt} &$\boldsymbol{-0.392}$\\
\multicolumn{2}{l}{YbF}  & & \\
      & \textbf{This work}  & \textbf{FSCC(0,1)+gaunt} & $\boldsymbol{-1.055}$\\
&\cite{Kozlov1995}Kozlov, 1995  & Semi-empirical from HFS & $-2.1$ \\
&  \cite{Titov1996}Titov, 1996 & RASSCF& $-1.3$  \\
 &\cite{Quiney1998}Quiney, 1998 & DHF  & $-0.6$  \\
&\cite{Quiney1998}Quiney, 1998 & DHF+CP &  $-1.3$  \\      
\multicolumn{2}{l}{YbOH}  & & \\
      & \textbf{This work}  & \textbf{FSCC(0,1)+gaunt}& $\boldsymbol{-1.067}$\\
      & \cite{Maison2019}Maison, 2019 & FSCC+gaunt & $-1.074$ \\
\hline\hline  
    \end{tabular*}

\end{table}

\section{Implications}

Like an EDM, a MQM will induce a molecular dipole moment that can be measured using spin-precession methods\cite{andreev2018improved,Cairncross2017,Hudson2011}.  However, unlike an eEDM, the induced molecular dipole moment will depend on the relative orientation of the nuclear and electron spins\cite{Sushkov1984,Flambaum2014,Skripnikov2015TaN} according to Equation 2.  This is important since the molecules under consideration, and generally those with nMQM sensitivity, have a non-zero effective electron spin $\boldsymbol{S'}$ and therefore eEDM sensitivity as well via the effective interaction $H_\text{eEDM}^\text{eff} = d_e \mathcal{E}_\text{eff}\Omega\propto\boldsymbol{S}'\cdot\boldsymbol{n}$ (see ref.\cite{Baron2017} for a comparison of the different conventions for effective eEDM interactions).  Because the molecules under consideration also have appreciable eEDM sensitivity\cite{denis2019,Maison2019}, this is a significant consideration for an experimental search.

Experimentally, this means that we can perform a search in states with different $G=I+S=I\pm1/2$, where $I$ is the nuclear spin of the heavy metal center, which is strongly coupled to the electron spin $S$ by the magnetic hyperfine interaction\cite{Wang2019}. ($F$ is often used in place of $G$ to indicate the coupled  metal nucleus and electron spin\cite{Sushkov1984,Skripnikov2015TaN}.)  These states are split in energy by $\sim A_\|$ and can therefore be resolved with lasers or microwaves.  By performing a molecular EDM spin-precession measurement in two such states, we can disentangle the nMQM contribution, which depends on $I$, and the eEDM contribution, which only depends on $S$.


One way in which nMQMs can arise is as a single-nucleon effect from a valence nucleon with a permanent EDM orbiting a nuclear core\cite{Dmitriev1994}.  nMQMs in non-spherical nuclei, specifically those possessing a quadrupole $(\beta_2)$ deformation, are significantly enhanced due to collective nuclear effects typically by a factor of\,\cite{Flambaum1994,Khriplovich1997,Flambaum2014} $\sim\beta_2 Z$.  These enhancements are directly proportional to the intrinsic nMQM $\mathcal{M}$ itself, as opposed to the MQM sensitivity parameter $W_M$, which comes primarily from electronic structure.  However, since the experimental observable is $\propto\mathcal{M}W_M$, enhancements to either quantity increase sensitivity to new $\mathcal{P,T}$-odd physics.  This means that Yb-containing compounds are over an order of magnitude more sensitive to $\mathcal{CP}$-violating hadronic physics, such as nucleon EDMs or strong $\mathcal{CP}$ violation\cite{Flambaum2014}, despite the fact that their sensitivity parameters $W_M$ differ by $\sim2.7$.  The quadrupole deformation parameter for $^{173}$Yb is\cite{Moller2016} $\beta_2\simeq 0.3$, which increases the size of the MQM by $\simeq 10$ compared to\cite{Flambaum2014,Moller2016} $^{137}$Ba ($\beta_2\simeq 0.05$).

Fundamentally, nMQMs arise due to $\mathcal{CP}$ violation in the hadronic sector, which is complex and contains several possible sources\cite{Liu2012,Engel2013}. MQMs therefore have overlap with searches for nuclear Schiff moments (NSMs), such as those using $^{199}$Hg~\cite{Graner2016}, $^{205}$TlF~\cite{Cho1991,Norrgard2017}, $^{225}$Ra~\cite{Parker2015,Bishof2016},  $^{129}$Xe~\cite{Allmendinger2019,Sachdeva2019} and $^{223}$Rn~\cite{Tardiff2008}.  Because of the complexity of the hadronic parameter space, all of these searches complement each other.  The most sensitive NSM measurement from the $^{199}$Hg atom\cite{Graner2016} lists 10 $\mathcal{CP}$-violating sources that are probed by the measurement, meaning that multiple searches in species with different sensitivities to the underlying sources are necessary to obtain robust bounds~\cite{Chupp2015,Flambaum2019Schiff}.  This need becomes even more apparent when remembering that the goal of such searches is to \emph{measure} $\mathcal{CP}$-violating moments, and a single positive measurement in a single system will not be able to isolate the source.

\section{Conclusion}
Throughout this study, an elaborate fully relativistic correlation approach, namely the coupled-cluster method, was employed to determine the values of the nuclear magnetic quadrupole moment interaction constant in systems currently of interest for the search of $\mathcal{CP}$ violation due to their sensitivity to various $\mathcal{P,T}$-odd interactions and their amendability to laser-cooling.
An extensive investigation of computational parameters was performed; the basis set, the treatment of relativity, the size of the virtual space and the treatment of correlation were scrutinized in order to optimize the model employed for the final values and estimate their uncertainty.
Thus we were able to report on the nMQM interaction constants with a $2\%$ error bar for BaF and BaOH and $4\%$ for YbF and YbOH.
Besides, the parallel magnetic hyperfine interaction constant and the electric field gradient were calculated to provide useful information for upcoming experiments as well as element of comparison with experimental data when available.

As previously highlighted in the study of the electron EDM interaction constants \cite{denis2019}, the isoelectronic fluorides and monohydroxides exhibit very similar enhancement of $\mathcal{CP}$-violating properties. This suggests that other ligands such as symmetric tops CH$_3$ or OCH$_3$ are very promising and might be worth investigating due to their unique experimental advantages \cite{Kozyryev2017}.

\section*{Acknowledgements}
The authors would like to thank the Center for Information Technology of the University of Groningen for providing access to the Peregrine high performance computing cluster and for their technical support.

\newpage
\bibliography{nMQM.bib}
\end{document}